%% file: BH_Ring.tex
\documentclass[prd,amssymb,intlimits,centertags,amsmath,twocolumn,showpacs,nofootinbib,floatfix]{revtex4}
\usepackage{graphicx}
\allowdisplaybreaks

\begin{document}

 \title{Black Holes Surrounded by Uniformly Rotating Rings}
 
 \author{Marcus Ansorg}
 \email{mans@aei.mpg.de}
 \affiliation{Max-Planck-Institut f\"ur Gravitationsphysik, Albert-Einstein-Institut, 14476 Golm, Germany}

 \author{David Petroff}
 \email{D.Petroff@tpi.uni-jena.de}
 \affiliation{Theoretisch-Physikalisches Institut, University of Jena,
           Max-Wien-Platz 1, 07743 Jena, Germany}

 \date{\today}

 \begin{abstract}
  \input Abstract
 \end{abstract}

 \pacs{04.70.Bw, 0.4.40.-b, 04.25.Dm \qquad preprint number: AEI-2005-103}
 
 \maketitle
 
 \section{Introduction}
  \input Intro

 \section{Field Equations and Boundary Conditions}
  \input Field_Eq
  
 \section{Physical Parameters}\label{physpar}
  \input Physpar
 \section{Numerical Methods}\label{methods}
  \input Methods

 \section{First Results}\label{results}
  \input Results

 \section{Future Work}\label{discussion}
  \input Future
 
 \begin{acknowledgments}
  We are grateful to R.\ Meinel for many interesting and helpful discussions. Thanks also to L.\ Rezzolla and A.\ Ashtekar  for bringing our attention to this topic. This work was supported in part by the Deutsche Forschungsgemeinschaft (DFG) through the SFB/TR7 ``Gravitationswellenastronomie''. 
 \end{acknowledgments}
  
 \bibliography{Reflink}
  
\end{document}

%% file: Abstract.tex
Highly accurate numerical solutions to the problem of Black Holes surrounded by uniformly rotating rings in axially symmetric, stationary spacetimes are presented. The numerical methods developed to handle the problem are discussed in some detail. Related Newtonian problems are described and numerical results provided, which show that configurations can reach an inner mass-shedding limit as the mass of the central object increases. Exemplary results for the full relativistic problem for rings of constant density are given and the deformation of the event horizon due to the presence of the ring is demonstrated. Finally, we provide an example of a system for which the angular momentum of the central Black Hole divided by the square of its mass exceeds one ($J_\text c/M_\text c^{\,2}>1$). 

%% file: Intro.tex
There are many reasons for choosing to study Black Holes with a surrounding ring. Both in the collapse of a single neutron star to a Black Hole and in the coalescence of two compact objects, it is expected that such a system exists, if only for a short time (see e.g.\ \cite{Shapiro95}, \cite{STU03}). The model considered here is also of interest for modelling massive Black Holes (and surroundings), which are now known to be contained in most galaxies. Furthermore, there exists speculation that the accretion of matter onto a Black Hole may be responsible for gamma ray bursts, e.g.\ \cite{RJ99}. In addition to this astrophysical motivation, there is interest in studying a Black Hole-ring system in order to see how matter affects the properties of the Black Hole. Finally, it seems worthwhile to study the few types of physical solutions to Einstein's equations that can be handled (even if only numerically) with extremely high accuracy. The numerical solutions that are obtained can also serve as initial data for a time evolution program. Because of the fact that very little is available in the way of good physical initial data for a two-body problem, such solutions are all the more important.

The problem of a slowly rotating Black Hole surrounded by an infinitesimal ring was handled perturbatively by Will in \cite{Will74,Will75}. The problem of accretion from a ring onto a central object and the importance of the self-gravitation of the ring was discussed by various authors in \cite{ACN83, Abramowicz84, BC92}. The dynamics of rings in the background metric of a Black Hole including the possibility of a runaway instability has been studied in \cite{RYZ03, MRY04, ZFRM05}. Lanza \cite{Lanza92} provided numerical solutions to the problem of an infinitely thin disc surrounding a Black Hole by using a multigrid method. Using an integral formulation of Einstein's equations, Nishida \& Eriguchi \cite{NE94} numerically solved the problem of a differentially rotating ring surrounding a Black Hole. They considered the ring to be a polytropic perfect fluid and prescribed a one-parameter rotation law. The methods used (more than ten years ago now) did not allow for an accuracy high enough to resolve the impact of the matter distribution on the Black Hole completely, and the authors were misled into making incorrect conjectures regarding the shape of Black Holes with zero angular momentum.

This paper is organised as follows. In \S~\ref{relativistic} we discuss Einstein's equations and the appropriate boundary conditions for describing the Black Hole and then turn our attention in \S~\ref{Newtonian} to related Newtonian problems. Section~\ref{physpar} is devoted to defining various physical quantities and \S~\ref{methods} to the numerical methods and some of the difficult issues that arise. We present first results for both the Newtonian and relativistic scenarios in \S~\ref{results}, providing examples for homogeneous rings and paying particular attention to the effect of the ring on the Black Hole. We recapitulate some of the results in \S~\ref{discussion} and discuss future plans.

%% file: Field_Eq.tex
\subsection{Relativistic Equations}\label{relativistic}

The equations and boundary conditions that hold for a stationary, axisymmetric, asymptotically flat spacetime containing a Black Hole and a fluid with purely rotational motions were discussed lucidly and at length in \cite{Bardeen73}. In this paper we adopt for the most part the notation used there and summarize the results that are relevant for this work. The line element can be written as
\[ ds^2 = -e^{2\nu}\,dt^2 + \varrho^2 B^2 e^{-2\nu} \left(d\varphi - \omega\, dt\right)^2
          + e^{2\mu}\left(d\varrho^2 + d\zeta^2 \right), \]
where the metric funtions $\nu$, $B$, $\omega$, and $\mu$ depend only on $\varrho$ and $\zeta$. For a region of spacetime in which the pressure $p$ is zero, it is possible to find a coordinate transformation for $\varrho$ and $\zeta$ yielding $B=1$. Since, however, our spacetime contains a ring with pressure, such a transformation cannot be performed globally. In the absence of a Black Hole, the requirement that the normal derivatives of the metric functions be continuous everywhere (even across the surface of the ring) together with the regularity of $B$ specifies the coordinates uniquely. When a Black Hole is present, however, there is a singularity inside the horizon, and regularity cannot be required everywhere. We thus use the additional coordinate freedom we have to choose coordinates in which the event horizon is a sphere and then excise the region inside the event horizon. Having chosen the horizon to be a sphere, it is natural to introduce the spherical coordinates $r$ and $\theta$ defined by
\[ \varrho = r \sin\theta, \ \ \  \zeta = r\cos\theta.\]
The location of the horizon will be denoted by
\[ r = \text{constant} =: r_{\text c}.\]

The fact that this two-surface is indeed an event horizon is realised by imposing the boundary conditions%
\footnote{The boundary condition for $B$ reads more generally $\varrho B=0$. In the coordinates (Weyl coordinates) for which $B=1$ holds, this implies that the event horizon must be a piece of the coordinate axis $\varrho=0$. \label{Weyl}}
\begin{align}\label{BC}
 \begin{split}
  e^{2\nu} = & \, 0 \\
  B        = & \, 0 \\
  \omega   = & \, \text{constant} =: \Omega_{\text c}.
 \end{split}
\end{align}

Whereas $\nu$ tends to $-\infty$ as one approaches the horizon, the quantity
\begin{equation}\label{eq:u} 
 u:= \nu - \ln B 
\end{equation}
is a regular function everywhere outside of the Black Hole, which makes it appropriate for numerical calculations.

On the horizon, it is possible to define a constant
\begin{equation}\label{kappa}
  \kappa := \left. e^{-\mu}\, \frac{\partial}{\partial r} e^{\nu}\right|_{r=r_{\text c}},
\end{equation}
which plays the role of temperature in Black Hole thermodynamics (see e.g.~\cite{Wald}).

The energy-momemtum tensor for the ring is taken to be that of a perfect fluid
\[ T^{ab} = (\varepsilon + p)\,u^a u^b + p\,g^{ab}, \]
where $\varepsilon$ is the energy density, $p$ the pressure and $u^a$ the four-velocity of a fluid element. Introducing the angular velocity of the matter in the ring relative to infinity $\Omega_{\text r}=d\varphi/dt$ and the velocity
\[ v:= \varrho B e^{-2\nu} (\Omega_{\text r}-\omega)\]
measured for a fluid element by a zero angular momentum observer, we can write the field equations as:
\begin{subequations}\label{field_equations}
 \begin{align}
  \begin{split}\label{eq:nu}
   & \nabla\cdot(B\nabla\nu)  - \frac{1}{2}\varrho^2B^3e^{-4\nu} (\nabla \omega)^2
	 =   \\
	  &\phantom{mm} 4\pi e^{2\mu}B\left[(\varepsilon+p)\frac{1+v^2}{1-v^2} + 2p \right] 
  \end{split} \\
  \begin{split}\label{eq:omega}
   &\nabla \cdot (\varrho^2 B^3 e^{-4\nu} \nabla \omega) = \\
	  &\phantom{mm} -16\pi \varrho B^2 e^{2\mu-2\nu}\, (\varepsilon+p)\, \frac{v}{1-v^2}
  \end{split} \\
  \begin{split}\label{eq:B}
   & \nabla \cdot (\varrho\nabla B) = 16\pi \varrho B e^{2\mu} p
  \end{split} \\
  \begin{split} \label{eq:mu}
   & \triangle_2\, \mu - \frac{1}{\varrho}\frac{\partial \nu}{\partial \varrho} 
    + \nabla\nu\nabla u - \frac{1}{4}\varrho^2 B^2e^{-4\nu}(\nabla\omega)^2 =\\
   & \qquad = -4\pi e^{2\mu}(\varepsilon+p).
  \end{split}  
 \end{align}
\end{subequations}
Here the operator $\nabla$ has the same meaning as in a Euclidean three-space in which $\varrho$, $\zeta$ and $\varphi$ are cylindrical coordinates. Thus the first three of the field equations can be applied as they are in $r$, $\theta$, $\varphi$ coordinates. In  eq.~\eqref{eq:mu}, the operator $\triangle_2:=\partial^2/\partial\varrho^2 + \partial^2/\partial\zeta^2$ is not coordinate independent.

As an alternative to eq.~\eqref{eq:mu}, one can combine the Einstein equations to arrive at two first order differential equations for $\mu$, the integrability condition of which is guaranteed to hold as a result of the Bianchi identities. Using this formulation, $\mu$ can be found via a line integral once $\nu$, $B$ and $\omega$ are known.

At the boundary of the ring, which is defined to be the surface of vanishing pressure, the following condition holds:
\begin{equation}\label{Ring-boundary}
 e^{2\nu}\,\left(1-v^2 \right) = \text{constant} =: e^{2 V_0},
\end{equation}
where $e^{2 V_0}$ is the value for -$g_{tt}$ in a frame of reference rotating together with the ring. 

By making use of the boundary conditions \eqref{BC} on the horizon and the field equations \eqref{eq:nu}--\eqref{eq:B}, we can derive the following further conditions that must hold on the horizon:
\begin{equation}\label{partial_BC}
 \frac{\partial^2 B}{\partial r^2} = -\frac{3}{r_{\text c}}\frac{\partial B}{\partial r}, \qquad
 \frac{\partial u}{\partial r}     = \frac{1}{r_{\text c}}, \qquad
 \frac{\partial \omega}{\partial r}     = 0.
\end{equation}

\subsection{Newtonian Equations}\label{Newtonian}

There are two reasons for our considering a Newtonian central body surrounded by a ring. On the one hand, it is generally helpful to consider a Newtonian problem before turning to a related one within the scope of general relativity. On the other hand, Newtonian theory will provide us with an approximative solution to Einstein's equations for the problem being considered here, which we require for our numerical methods.

The first thought that comes to mind when looking for an approximative solution to Einstein's equations, is to use the analytically known Schwarzschild (or Kerr) solution surrounded by a test-ring.%
\footnote{By `test-ring' we mean a ring without self-gravitation.}
We know from Newtonian theory however, that a central object surrounded by a uniformly rotating test-ring of finite dimension cannot remain in equilibrium.

To see this, consider the accelerations of two fluid elements in the equatorial plane of the ring, one at the inner and the other at the outer edge. In a corotating frame of reference, the accelerations have three sources: the gravitational attraction to the central object, the pressure gradient within the ring and the centrifugal effects. Remembering that the ring does not influence the gravitational field of the central object, it is clear that the field strength at the location of the inner particle is greater than that at the outer one. The pressure gradient causes an acceleration acting toward the coordinate origin at the inside of the ring and toward infinity at the outside. Since the ring is taken to be in uniform rotation, the centrifugal acceleration must be greater at the outer edge than at the inner one. Now each of these three accelerations tends to increase the separation between the two particles so that their sum must rip the ring apart.

Since we therefore do not have a relativistic solution at hand, we turn to Newtonian theory. The Newtonian potential for a uniformly rotating ring of constant density surrounding a central body is, of course, a solution of the Poisson equation
\begin{equation}\label{Poisson} 
 \triangle U = 4\pi\varepsilon,
\end{equation}
where $\varepsilon=\varepsilon_{\text c}+\varepsilon_{\text r}$, $\varepsilon_{\text c}$ being the source of the central body and $\varepsilon_{\text r}$ the mass density, here taken to be constant, inside the ring and zero elsewhere. Because the potential due to the central object is often singular, it is convenient for numerical reasons to work with the potential of the ring alone
\[ U_{\text r} :=U-U_{\text c}, \]
where $U_{\text c}$ is the part of the potential arising from $\varepsilon_{\text c}$. Consider, for example the situation in which the central body is a point particle. Then the potential 
\[ U_{\text r}=U-M_{\text c}/r, \]
where $M_{\text c}$ is the mass of the central object, is regular everywhere and is thus better suited to numerical calculations than $U$. On the boundary of the ring, $U_{\text r}$ must obey the equation
\begin{equation}\label{boundary}
 U_{\text r} = \frac{1}{2}\left(\Omega_{\text r}\,r\sin\theta\right)^2 +
  V_0 + M_{\text c}/r,
\end{equation}
where $V_0$ is the constant ``corotating potential''.

In order to prepare the groundwork for the approximative solution to the Einstein equations that was discussed at the beginning of this section, we now consider the situation in which the central potential $U_{\text c}$ takes on a different form. As was mentioned in footnote~\ref{Weyl}, a Black Hole in Weyl coordinates is located along the axis of symmetry. In analogy, we now consider a line of mass of constant linear mass density located along the axis of rotation $\varrho_{\text W}=0$ and extending from $\zeta_{\text W}=-M_{\text c}$ to $\zeta_{\text W}=M_{\text c}$, where $(\varrho_{\text W},\zeta_{\text W},\varphi_{\text W})$ are cylindrical coordinates and $M_{\text c}$ is the total mass of the infinitely thin rod. The potential for such a configuration is given by
\begin{align}\label{U_Weyl}
 U_{\text c} &= \, -\frac{1}{2} \int\limits_{-M_{\text c}}^{M_{\text c}} \frac{dz}{\sqrt{{\varrho_{\text W}}^2+\left(\zeta_{\text W}-z\right)^2}} \nonumber \\
   &= \, -\frac{1}{2} \ln\left(\frac{ M_{\text c}-\zeta_{\text W} + \sqrt{{\varrho_{\text W}}^2 + \left(M_{\text c}- \zeta_{\text W}\right)^2 }  }
	                                 { -M_{\text c}-\zeta_{\text W} + \sqrt{{\varrho_{\text W}}^2 + \left(M_{\text c}+ \zeta_{\text W}\right)^2 }  } \right). 
\end{align}
If we introduce the coordinates $(\varrho,\zeta,\varphi)$ defined by
\begin{equation*}
 \begin{aligned}
  \varrho_{\text W} &= \, \varrho\left(1-\left(\frac{M_{\text c}}{2r}\right)^2 \right) \\
  \zeta_{\text W} &= \, \zeta\left(1+\left(\frac{M_{\text c}}{2r}\right)^2 \right) \\
  \varphi_{\text W} &=\, \varphi
 \end{aligned}
\end{equation*}
with $r:=\sqrt{\rho^2+\zeta^2}$, then the original line of mass becomes a sphere of radius $r=M_{\text c}/2$ and eq.~\eqref{U_Weyl} becomes
\begin{equation*}
 U_{\text c} = \ln\left(\frac{1-M_{\text c}/2r}{1+M_{\text c}/2r}\right).
\end{equation*}
Defining the quantities
\begin{align} 
 B:=&\, 1-\left(M_{\text c}/2r\right)^2 \label{B_Newton} \intertext{and}
 u_{\text c}:=&\, U_{\text c} -\ln B = -2\ln\left({1+M_{\text c}/2r}\right),
  \label{Utilde}
\end{align}
we find that at the radius $r=M_{\text c}/2$, the following conditions hold: $e^{2U_{\text c}}=0$, $B=0$ and $u_{\text c}$ is a regular function. Furthermore, $B$ is a solution to eq.~\eqref{eq:B} in the vacuum region. Because of this complete analogy to the relativistic case (see eqs~\eqref{BC} and \eqref{eq:u}), and because $\omega$ is small in comparison to the other potentials for small total mass, we can use the potentials described above to construct an initial solution for the numerical program as will be discussed in more detail in \S~\ref{methods}. 

%% file: Physpar.tex
As one approaches spatial infinity, the metric functions behave as
\begin{equation}\label{asymptotics}
 \begin{alignedat}{4}
  \nu     &= \frac{-M_{\text{tot}}}{r}   \,&+&\, {\cal O}\left(\frac{1}{r^2}\right)    
                  & \qquad B &= \quad 1 \,&+&\, {\cal O}\left(\frac{1}{r^2}\right) \\
  \omega  &= \frac{2J_{\text{tot}}}{r^3} \,&+&\, {\cal O}\left(\frac{1}{r^4}\right)  
                  & \qquad \mu &= \frac{M_{\text{tot}}}{r} \,&+&\, {\cal O}\left(\frac{1}{r^2}\right). 
 \end{alignedat}
\end{equation}
Since the spectral methods used here involve compactifying all of spacetime onto various domains, the above equations can be used to read off the total mass and total angular momentum directly from infinity.

The total angular momentum, as well as the individual angular momenta of the ring and the Black Hole, can also be found by integrating eq.~\eqref{eq:omega}. The integral of the right hand side, which clearly vanishes in the vacuum region, is a multiple of the angular momentum of the ring, namely  $-16\pi\,J_{\text r}$ (see \cite{Bardeen73} for more details). Thus we have
\begin{equation}\label{int_Jring}
 J_{\text r} = \iiint  e^{2\mu-2\nu} \varrho B^2 (\varepsilon+p) \frac{v}{1-v^2} \, \varrho\,d\varrho\, d\zeta\, d\varphi,
\end{equation}
where the integral is performed over the entire matter region. The integral on the left hand side of eq.~\eqref{eq:omega} can be converted using the divergence theorem into a surface integral at infinity and one at the horizon. Using the asymptotic behaviour given in eq.~\eqref{asymptotics}, the surface integral at infinity can be shown to be equal to $-16\pi\,J_{\text{tot}}$. The equality of eq.~\eqref{eq:omega} itself means that the surface integral over the horizon yields the angular momentum of the Black Hole $J_{\text c}$. We thus have
\begin{align}\label{JBH}
 J_{\text c} &=\, \frac{-1}{16\pi} \iint \left. r^2 \sin^2\theta\, B^3 e^{-4\nu} \frac{\partial \omega}{\partial r}
                    r^2\,\sin\theta \, d\theta\,d\varphi\  \right|_{r=r_{\text c}} \nonumber \\
             &=\, \frac{-r_{\text c}^{\,4}}{8} \int_0^\pi \left. \sin^3\theta\, e^{-4u} \frac{1}{B} \frac{\partial \omega}{\partial r}
                   d\theta\  \right|_{r=r_{\text c}} \nonumber \\
             &=\, \frac{-r_{\text c}^{\,4}}{8} \int_0^\pi \left. \sin^3\theta\, e^{-4u} \frac{\partial^2 \omega}
                       {{\partial r}^2}\left(\frac{\partial B}{\partial r}\right)^{-1}d\theta\  \right|_{r=r_{\text c}},
\end{align}
where l'H\^opital's rule was used to get from the second to the third line ($\left.\partial \omega/\partial r\right|_{r=r_{\text c}}=0$, cf.\ eq.~\eqref{partial_BC}).

We can proceed similarly in order to calculate the components of mass. Taking the combination of equations $\eqref{eq:nu}-\omega\eqref{eq:omega}/2$, we obtain from the integral over the matter distribution
\begin{align}\label{int_Mring}
 \begin{split}
 M_{\text r} &=\, \iiint   e^{2\mu} B \biggl[ (\varepsilon+p) \frac{1+v^2}{1-v^2} + 2p   \\
            & \phantom{\, \iiint}            + 2\varrho B e^{-2\nu}(\varepsilon+p)\,\omega\frac{v}{1-v^2}\biggr]
                          \, \varrho\,d\varrho\, d\zeta\, d\varphi
 \end{split} \nonumber \\
 \begin{split}
             &=\, \iiint   e^{2\mu} B \Biggl[ (\varepsilon+p) \frac{1+v^2}{1-v^2} + 2p\, +  \\
            & \qquad + 2 \varrho B e^{-2\nu}(\varepsilon+p)\,(\omega-\Omega_{\text r})\frac{v}{1-v^2}\, + \\
            & \qquad + 2 \varrho B e^{-2\nu}(\varepsilon+p)\,\Omega_{\text r}\,\frac{v}{1-v^2}\Biggr]
                          \, \varrho\,d\varrho\, d\zeta\, d\varphi
 \end{split} \nonumber \\
 \begin{split}
             &=\, \iiint   e^{2\mu} B \Biggl[ (\varepsilon+p) \frac{1+v^2}{1-v^2} + 2p\, +  \\
            & \qquad + 2 (\varepsilon+p)\frac{v^2}{1-v^2}\, \Biggr]
                          \, \varrho\,d\varrho\, d\zeta\, d\varphi \, + \\
            & \qquad + 2 \Omega_{\text r}\,J_{\text r}
 \end{split} \nonumber \\
  \begin{split}
             &=\, \iiint   e^{2\mu} B (\varepsilon+3p) \, \varrho\,d\varrho\, d\zeta\, d\varphi \, + \\
            & \qquad + 2 \Omega_{\text r}\,J_{\text r}.
 \end{split} 
\end{align}
Note that in the step from the second to the third equals sign, $\Omega_{\text r}$ was pulled out of the integral, which is only valid for uniform rotation. The left hand side of the equation can again be written as a total divergence and one finds $4\pi M_{\text{tot}}$ for the surface integral at infinity. The surface integral over the horizon yields
\begin{align}\label{MBH}
 \begin{split}
 M_{\text c} &=\, \frac{r_{\text c}^{\,2}}{4\pi} \iint \biggl[ B\,\frac{\partial \nu}{\partial r} - \\
             &\left. \qquad
              -\frac{1}{2}r_{\text c}^{\,2} \sin^2\theta B^3 e^{-4\nu}\omega\frac{\partial \omega}{\partial r} \biggr]
                        \,\sin\theta \, d\theta\,d\varphi\  \right|_{r=r_{\text c}} \nonumber
 \end{split}\\
            &= \frac{r_{\text c}^{\,2}}{2} \int_0^\pi \left.  B\,\frac{\partial \nu}{\partial r} 
                        \,\sin\theta \,d\theta\  \right|_{r=r_{\text c}} 
                         + 2 \Omega_{\text c} J_{\text c} \nonumber \\
            &= \frac{r_{\text c}^{\,2}}{2} \int_0^\pi \left. \frac{\partial B}{\partial r} 
                        \,\sin\theta \,d\theta\  \right|_{r=r_{\text c}} 
                         + 2 \Omega_{\text c} J_{\text c}.
\end{align}

The last two quantities we wish to define are the proper equatorial and polar radii (the former is often called the circumferential radius). They are defined by taking the invariant length of the closed loop along the horizon with $d\theta=0$ and $d\varphi=0$ respectively, and dividing by $2\pi$. The proper equatorial radius is
\begin{equation}
 R_{\text e} = r_{\text c}\, e^{-u(r=r_{\text c},\theta=\pi/2)}.
\end{equation}
For $R_{\text p}$ we find%
%
%
\begin{align}
 R_{\text p} &= \frac{r_{\text c}}{\pi} \int_0^\pi \left. e^{\mu} \right|_{r=r_{\text c}} \, d\theta \nonumber \\
             &= \frac{r_{\text c}}{\pi \kappa} \int_0^\pi \left. e^{u}\, \frac{\partial B}{\partial r}
                          \right|_{r=r_{\text c}} \, d\theta,
\end{align}
where $\kappa$ is defined in Eq.~\eqref{kappa}.

\begin{figure}
 \includegraphics[width=\columnwidth]{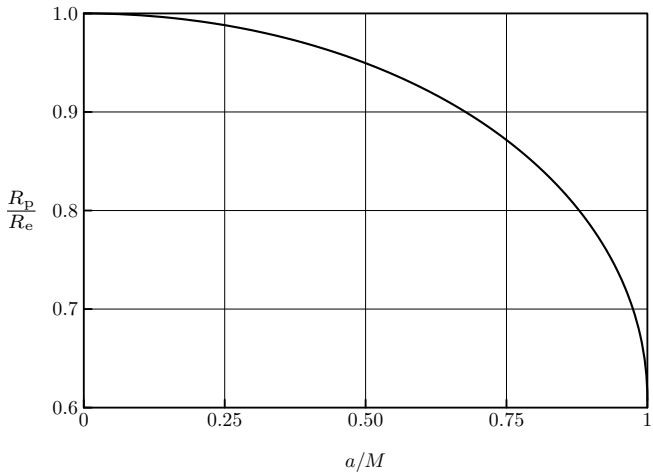}
 \caption{The value of the radius ratio $R_{\text p}/R_{\text e}$ for a Kerr Black Hole as the parameter $a/M$ is varied from 0 to 1.                 \label{Kerr_radius_ratio}}
\end{figure}

For the purposes of later comparison, we also write down the ratio of proper polar to equatorial radius for the Kerr metric:
\begin{equation}\label{eq:Kerr_RR}
 \left(\frac{R_{\text p}}{R_{\text e}}\right)_{\text {Kerr}}= \frac{\sqrt{2 r_+}}{\pi\sqrt{M}}\,E\left(\frac{a}{\sqrt{2M\,r_+}}\right),
\end{equation}
where $a=J/M$, $r_+=M+\sqrt{M^2-a^2}$, $J$ and $M$ are the angular momentum and mass of the Kerr Black Hole and $E(k)$ is the complete elliptic integral of the second kind of modulus $k$. The value of $R_{\text p}/R_{\text e}$ falls monotonically from 1 to $0.60800\ldots$ as can be seen in Fig.~\ref{Kerr_radius_ratio}.

%% file: Methods.tex
In order to solve both the Newtonian and the relativistic free-boundary problem, we resort to a multi-domain, pseudo-spectral method. The techniques used are in essence those described in \cite{AKM3} and we here provide only a brief description of the general method, concentrating however on those issues that are unique to the problem being considered here. We shall describe the method used to solve the relativistic problem of a ring circumscribing a Black Hole (i.e.\ eqs~(\ref{field_equations}) together with the appropriate boundary conditions, asymptotic behaviour and regularity conditions), but the method can be applied with minor modifications to the Newtonian problems discussed above.

As was mentioned in \S~\ref{relativistic}, the metric functions depend only on $\varrho$ and $\zeta$. Assuming reflectional symmetry with respect to the equatorial plane ($\zeta=0$), we need only consider half of the $\varrho$-$\zeta$ plane. The quarter circle located at $\varrho^2+\zeta^2={r_{\text c}^{\,2}}$ is removed (i.e.\ we excise the interior of the Black Hole) and the remainder is divided up into five domains. It is essential that one of the domain boundaries coincide with the (unknown) surface of the ring. The location of the other three domain boundaries and indeed the number of domains that is chosen is somewhat arbitrary and we here describe a choice that has proved fruitful. In order to compactify the vacuum domain, we introduce the complex coordinate $\tilde{z}$, defined by
\begin{equation}
 z =: i \varrho_{\text m} \cot\frac{\tilde{z}}{2} \quad (z:=\varrho+i\zeta,\ \tilde z:=\tilde \varrho+i \tilde \zeta),
\end{equation}
whereby $\varrho_{\text m}$, which can take on any value between the inner radius $\varrho_{\text i}$ and outer radius of the ring $\varrho_{\text o}$, is here taken to be the arithmetic mean
\[ \varrho_{\text m}:= \frac{1}{2}\left(\varrho_{\text i}+\varrho_{\text o} \right).\]
Regularity along the axis and in the equatorial plane is ensured by introducing the coordinates
\begin{align}
 \begin{alignedat}{2}
  x         & := \varrho^2                      & \qquad y        &:= \zeta^2 \\
  \tilde x  & := \sin^2 \frac{\tilde\varrho}{2} & \qquad \tilde y &:= \sinh^2\frac{\tilde \zeta}{2}
 \end{alignedat}
\end{align}
and requiring that the potentials in these coordinates be analytic there. A simple calculation shows that the relation between these coordinates can be expressed as
\begin{align}
 \begin{split}\label{xy_of_xsys}
  &x =  \frac{x_{\text m}\tilde y\,(1+\tilde y)}{(\tilde x + \tilde y)^2},\qquad
  y =  \frac{x_{\text m}\tilde x\,(1-\tilde x)}{(\tilde x + \tilde y)^2}
 \end{split} 
\intertext{with}
  &x_{\text m} := \varrho_{\text m}^{\,2}. \nonumber
\end{align}    
It is clear from eq.~\eqref{xy_of_xsys} that $\tilde x$ is defined on the interval $[0,1]$. As can be seen in Fig.~\ref{xs-ys}, $\tilde y$ is bounded in the vacuum region by 0 from below and by the surface of the ring from above.

\begin{figure}
 \includegraphics{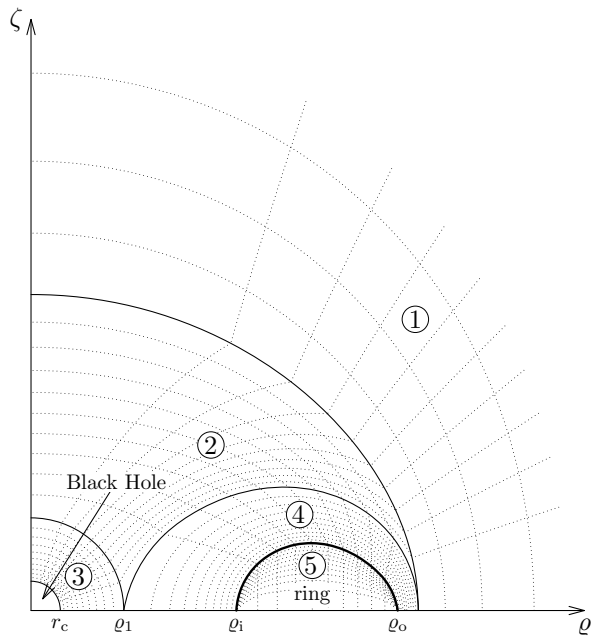}
 \caption{The division of the $\varrho$-$\zeta$ plane into the domains used in the spectral methods. The fifth domain is the interior of the ring and the black hole is at the position of the quarter circle that is extracted from the origin. The physical parameters chosen in this example are $\varrho_{\text i}/\varrho_{\text o}=0.56$ and $r_{\text c}/\varrho_{\text o}=0.08$. The further domain divisions were chosen by setting $\tilde x_0=0.45$, $\tilde y_0=1.2$ and $x_1=\varrho_1^{\,2}=0.064\varrho_\text o^{\,2}$, which implies that domain 3 is a spherical shell with an inner radius of $0.08\varrho_\text o$ and an outer radius of $\approx 0.2530\varrho_\text o$. \label{rho-zeta}}
\end{figure}

\begin{figure}
 \includegraphics{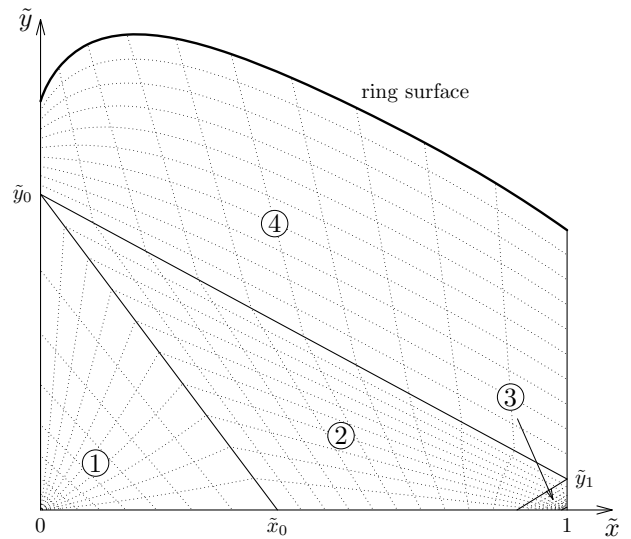}
 \caption{The vacuum region of Fig.~\ref{rho-zeta} depicted in $\tilde x$-$\tilde y$ space. Note that the plot is scaled such that two units in the $\tilde x$ direction correspond to one unit in the $\tilde y$ direction. \label{xs-ys}}
\end{figure}

An exemplary division of the $\varrho$-$\zeta$ plane into five domains can be found in Fig.~\ref{rho-zeta}.
Each of the five domains is mapped onto the square $(s,t) \in I^2=[0,1]\times[0,1]$ as follows:
\begin{subequations}\label{coord}
 \begin{align}
  \begin{split}\label{coord1}
   \text{domain 1:}\quad \tilde x & = \tilde x_0\, s^2\,(1-t) \\
                    \tilde y & = s^2\,t\,\left(s\,(\tilde y_0-\tilde x_0) + \tilde x_0\right)
  \end{split} \\   \nonumber \\
  \begin{split}\label{coord2}
   \text{domain 2:}\quad \tilde x & = s\,(1-t)\,\tilde x_0 + \frac{(1-s)(x_{\text m}-x_1 t)}{x_{\text m}+x_1(1-2t)} \\
                         \tilde y & = s\,t\,\tilde y_0 + \frac{(1-s)\,x_1 t}{x_{\text m}+x_1(1-2t)}
  \end{split} \\ \nonumber \\
  \begin{split}\label{coord3}
   \text{domain 3:}\quad \tilde x & = 1 + \frac{r_{\text c}^{\,2}(t-1)}{x_{\text m}\sigma+r_{\text c}^{\,2}(1-2t)} \\
                         \tilde y & = \frac{r_{\text c}^{\,2}t}{x_{\text m}\sigma+r_{\text c}^{\,2}(1-2t)} \\
   \text{with}\quad \sigma &:=\left(\frac{r_{\text c}^{\,2}}{x_1}\right)^{1-s}
  \end{split} \\ \nonumber \\
  \begin{split}\label{coord4}
   \text{domain 4:}\quad \tilde x & = t\,(1-s) + (1-t)\,\tilde x_{\text s}(s) \\
                   \quad \tilde y & = t\,(s\,\tilde y_0 + (1-s)\,\tilde y_1) + (1-t)\,\tilde y_{\text s}(s)  
  \end{split} \\ \nonumber \\
  \begin{split}\label{coord5}
   \text{domain 5:}\quad       x & = \varrho^{\,2}_{\text i} + s\,(\varrho^{\,2}_{\text o}-\varrho^{\,2}_{\text i}) \\
	                \quad       y & = (1-t)\,y_{\text s}(s). 
  \end{split}
 \end{align}
\end{subequations}

\begin{figure}
 \includegraphics{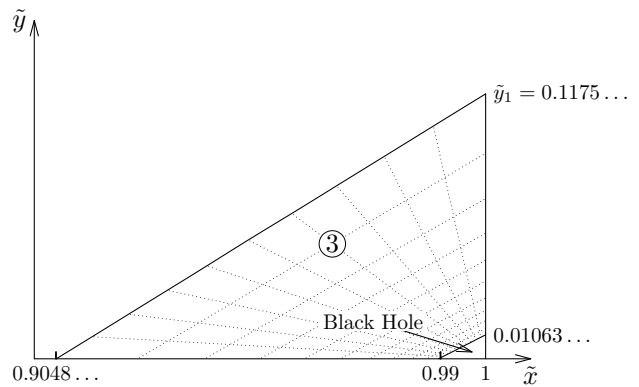}
 \caption{A blowup of domain 3 of Fig.~\ref{xs-ys}. \label{xs-ys_big}}
\end{figure}

The meaning of the various quantities not yet defined  can be explained most easily by referring to Figs~\ref{rho-zeta}, \ref{xs-ys} and \ref{xs-ys_big}. The constants $\tilde x_0$, $\tilde y_0$, and $x_1=\varrho_1^{\,2}$ are chosen, as appropriate, for the configuration being considered, and can be seen in Figs~\ref{xs-ys} and \ref{xs-ys_big}. It then follows from  eq.~\eqref{xy_of_xsys} that $\tilde y_1=x_1/(x_\text m - x_1)$. We choose the domain boundary between domains 2 and 3 to be a circle in $\varrho$-$\zeta$ coordinates (see Fig.~\ref{rho-zeta}). The surface of the ring, which must be solved for as part of the global problem, enters into the coordinate transformation via $y_{\text s}(s)$ of eq.~\eqref{coord5}. The value for $(\tilde x_{\text s}(s),\tilde y_{\text s}(s))$ can be found once $y_{\text s}(s)$ is known by inverting eq.~\eqref{xy_of_xsys} and taking
$x_{\text s}(s)=\varrho^{\,2}_{\text i} + s\,(\varrho^{\,2}_{\text o}-\varrho^{\,2}_{\text i})$.

Each of the metric potentials as well as the function describing the boundary of the ring is expanded in terms of Chebyshev polynomials and truncated at a predetermined order. The boundary conditions at the event horizon, the asymptotic behaviour and the continuity of the functions at the domain boundaries is guaranteed by the specific representation for the potentials that is employed in the program (see \cite{AKM3} for more details). What remains is to prescribe four physical parameters (for example the mass and angular momentum for each of the two objects) and formulate $n$ equations to solve for the $n$ unknown coefficients in the polynomial representation of the functions. We formulate the Einstein equations at the collocation points%
\footnote{For an introduction to spectral methods including a definition of collocation points, see e.g.\ \cite{Spectral}.}
in the interior of each domain and require that the normal derivatives of the metric functions be continuous at the collocation points along the one-dimensional domain boundaries. This leads to an algebraic system of non-linear equations for the Chebyshev coefficients, which is then solved using a Newton-Raphson method.

The Newton-Raphson method relies on an initial ``guessed'' solution that cannot be far away from the desired solution if the method is to converge. Most of the time, we simply take an existent solution as the inital guess and vary the four parameters in order to arrive at a new solution. The question arises, however, as to how one goes about constructing the very first solution. Ideally, one would like to have an analytic solution as a limiting solution to the problem being studied, and could use such a solution as an initial ``guess''. When dealing with one-body problems, such analytic solutions are available: the Maclaurin spheroids in the Newtonian limit, the global Schwarzschild solution in the static limit or the relativistic disc of dust \cite{NM95} in the highly flattened limit, for example. As was shown in \S~\ref{Newtonian}, the Newtonian test-ring limit does not exist if one restricts oneself to uniform rotation however. Thus, we do not expect the limit of a Schwarzschild (or Kerr) Black Hole surrounded by a test-ring to exist either. We do expect, however, that it  will be possible to construct a sufficiently good initial guess by solving the Newtonian problem described in \S~\ref{Newtonian} of a ring surrounding a line of mass represented in coordinates in which the central mass is a sphere. This expectation relies on the fact that in the limit in which this sphere as well as the total mass become infinitesimal, one arrives at the point mass Newtonian limit of the relativistic situation.
 
To solve the Newtonian problem, we took the numerically determined potential of a ring (without a central body) from the program described in \cite{AKM2} and linearly superposed the potential $u_{\text c}$ of eq.~\eqref{Utilde} with $M_{\text c}/\varrho_{\text o} \ll 1$ in order to acquire an initial solution with which to solve the Newtonian two-body problem (where the potential on the boundary of the ring is given by Eq.~\eqref{boundary}). We then increased the value of $M_{\text c}/\varrho_{\text o}$ until the masses of the ring and the central object were comparable, but keeping the total mass small. This choice was made since the limit of a vanishingly small central mass is numerically difficult to handle for reasons that will be discussed in the next paragraph. Using $u=u_{\text c}+u_{\text r}$ as supplied by this program, $B$ given by Eq.~\eqref{B_Newton} and setting $\omega=0$, we created a successful initial file for ``starting up'' the relativistic program. 

An aspect of the two-body problem that presents some difficulty is the fact that two different length scales are of significance.  This is particularly pronounced as one approaches the limit in which the central Black Hole vanishes or in the weak relativistic regime when the total mass is small (in this limit, the sphere representing the Black Hole also becomes small). One can well imagine that if the mass of the Black Hole is significantly smaller than that of the ring, then the behaviour of the metric potentials throughout most of spacetime is essentially governed by the ring and will not differ significiantly from the behaviour that would be found were the Black Hole not there at all. Nonetheless, the fact that one can prescribe boundary values for $\omega$ on the horizon and the fact that $e^{2\nu}$ and $B$ {\it must} vanish there, means that the values of the metric functions very close to the Black Hole (i.e.\ somewhere in domain 3 of Figs.~\ref{rho-zeta}--\ref{xs-ys_big}) do differ significantly from their values elsewhere. As the mass of the Black Hole  grows smaller, the metric functions come closer and closer to being non-differentiable. This can lead to problems when trying to represent such functions using a Chebyshev expansion.

\begin{figure}
\includegraphics[width=\columnwidth]{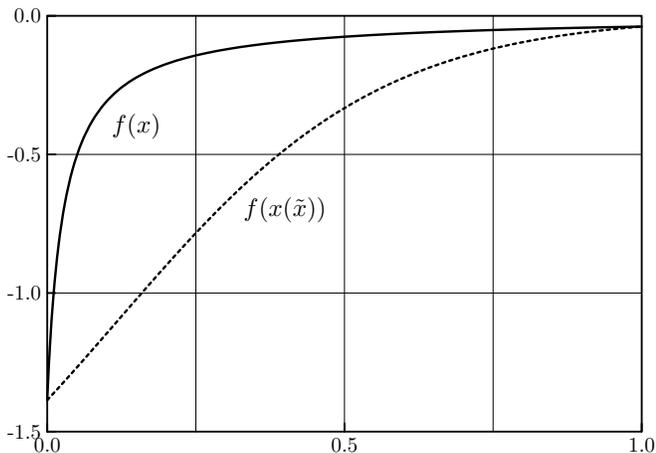}
\caption{The functions $f(x)$ (solid line) and $f(x(\tilde x))$ (dotted line) with $\delta=0.01$. See text for the definition of these two functions. \label{cheb_func}}
\end{figure}

\begin{figure}
\includegraphics[width=\columnwidth]{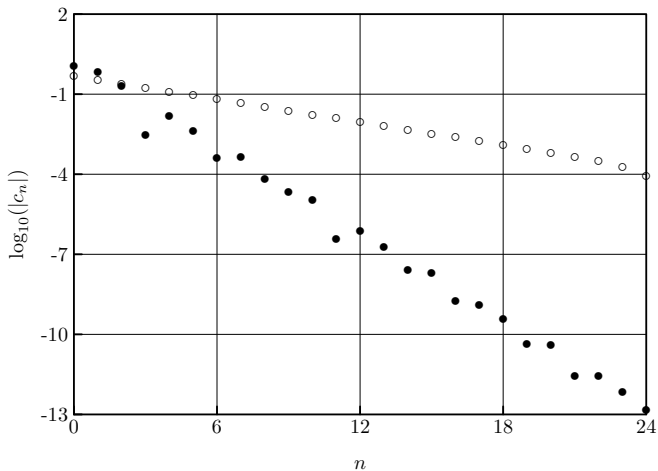}
\caption{The common logarithm of the absolute value of the Chebyshev coefficients $c_n$ is shown for an expansion of the functions from Fig.~\ref{cheb_func}. The expansions are of the form $f(x)\approx \frac{1}{2}\,c_0 + \sum_{n=1}^{24}\, c_n\, T_n(2x-1)$ and $f(x(\tilde x))\approx \frac{1}{2}\,\tilde c_0 + \sum_{n=1}^{24}\, \tilde c_n\, T_n(2\tilde x-1)$, where $T_n$ are Chebyshev polynomials of the first kind. The coefficients are determined such that the function and its polynomial approximation have the same value at each of the collocation points. The open circles show the coefficients $c_n$ and the filled circles $\tilde c_n$. \label{cheb_coeff}}
\end{figure}

The nature of this problem and the solution that we provide to it will now be demonstrated using exemplary functions with essentially the same behaviour as that of the metric potentials. The relativistic potential $u$ of eq.~\eqref{eq:u} for a Black Hole is similar in its qualitative behaviour to the Newtonian $u_{\text c}$ of eq.~\eqref{Utilde}. The potential of the ring, which we could model by an infinitesimal ring of constant density, is roughly constant in the vicinity of the central object. Consider therefore simply the function
\begin{align}\label{eq:f}
 f(x) =&\, 2\,\ln\left(\frac{1}{1+a}\right)
 \intertext{with}
 a =&\, \frac{2\delta}{x+2\delta}, \ \ x\in[0,1], \nonumber
\end{align}
where $\delta$ is a dimensionless mass parameter reflecting the size of the central object (for our relativistic two-body system, we can take e.g.\ $\delta=M_\text c/\varrho_\text o$). If $\delta$ is small, then $a$ is close to 0 everywhere except when $x$ approaches zero, since $a(0)=1$ holds for any value of $\delta$. Such a function cannot be approximated well using a Chebyshev expansion since the derivative of the function at the point $x=0$,
\[\left.\frac{df}{dx}\right|_{x=0} =\frac{1}{2\delta}, \]
is quite large for small $\delta$. Taking into account this behaviour, however, we can dramatically attenuate the problem by introducing an appropriately rescaled $x$
\begin{equation}\label{eq:ftilde}
 x = \delta\left[\left(1+\frac{1}{\delta} \right)^{\tilde x} -1 \right].
\end{equation}
The derivative of $f$ with respect to $\tilde x$ at $\tilde x=0$ is
\[\left.\frac{df}{d\tilde x}\right|_{\tilde x=0} =\frac{1}{2}\ln\left(1+\frac{1}{\delta} \right), \]
which merely grows logarithmically as $\delta$ tends to zero. A similar issue is encountered in the context of excision initial data for binary Black Holes with extreme radius ratios and is discussed in \cite{Ansorg05}. A comparison of the behaviour of the functions $f$ as it depends on $x$ and $\tilde x$ for $\delta=0.01$ can be found in Fig.~\ref{cheb_func}.
To see how well the Chebyshev expansion approximates a given function, one can look to see how quickly the coefficients grow small. Figure~\ref{cheb_coeff} shows the logarithm of the absolute value of the coefficients for the expansion of $f(x)$ and $f(x(\tilde x))$. As in the computer program described above, the coefficients were calculated by requiring that the polynomial expansion take on the value of the underlying function at the collocation points. One can see clearly that the coefficients of $f(x)$ do not show the rapid fall-off exhibited by those of $f(x(\tilde x))$. This is precisely the reason why we chose a coordinate transformation in domain~3, eq.~\eqref{coord3}, in which there is an exponential dependence on $s$. We shall see in \S~\ref{results} that we are indeed able to reach very small values for the mass of the Black Hole.

%% file: Results.tex
\subsection{A Newtonian Ring Surrounding a Point Mass}

\begin{figure}
 \includegraphics{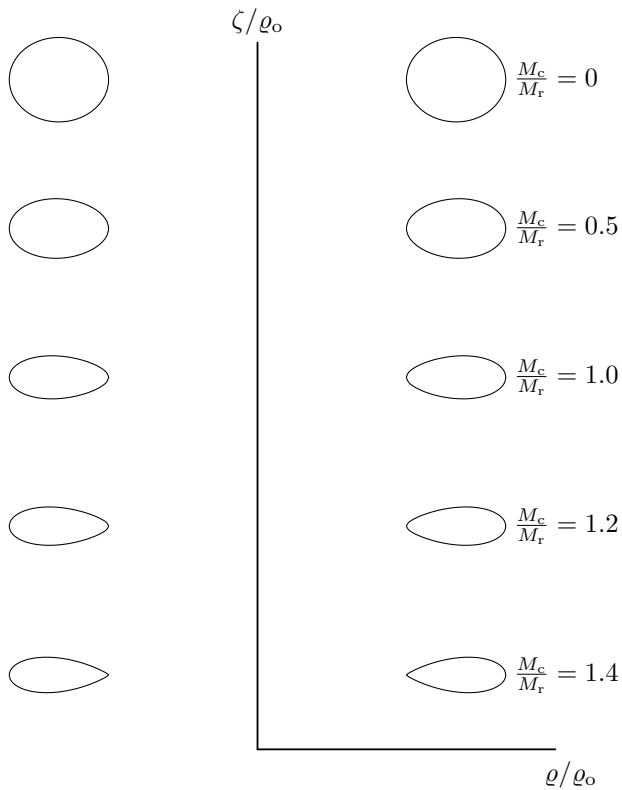}
 \caption{Cross-sections of Newtonian rings surrounding a point mass with varying ratios of central to ring mass $M_{\text c}/M_{\text r}$. The normalized coordinate $\zeta/\varrho_{\text o}$ is plotted agianst $\varrho/\varrho_{\text o}$. For each of these configurations, the ratio of inner to outer radius of the ring was chosen to be $\varrho_{\text i}/\varrho_{\text o}=0.6$ and for the normalized total mass we took  $M_{\text{tot}}\sqrt{\varepsilon}=(M_{\text c}+M_{\text r})\sqrt{\varepsilon}=1$. \label{Newt:seq}}
\end{figure}

In this subsection, we present a few characteristic features of the point mass-ring system. As was shown in \S~\ref{Newtonian}, a Newtonian, rigidly rotating test-ring of finite size cannot exist in equlibrium. Since we know that rings without a central body exist within Newtonian theory \cite{Poincare, Kowalewsky, Dyson92, Dyson93, Wong74, Eriguchi81, AKM2}, there must exist a maximum for the ratio of the mass of the central body to that of the ring if the ring is to remain a finite size. One would expect the gravitational pull towards the central object to grow ever stronger until mass-shedding at the inner edge sets in, i.e.\ until the gradient of pressure at the inner edge of the ring in the equatorial plane vanishes and a cusp develops, marking the point at which a fluid element is about to be pulled away from the ring. This is indeed what is observed. In Fig.~\ref{Newt:seq}, a sequence of rings about a point mass is shown for an increasing ratio of the central to the ring mass $M_{\text c}/M_{\text r}$. The ratio of inner to outer radius of the ring was held constant at the value $\varrho_{\text i}/\varrho_{\text o}=0.6$ and the total (normalized) mass of the system was taken to be $M_{\text{tot}}\sqrt{\varepsilon}=1$.

\begin{figure}
 \includegraphics{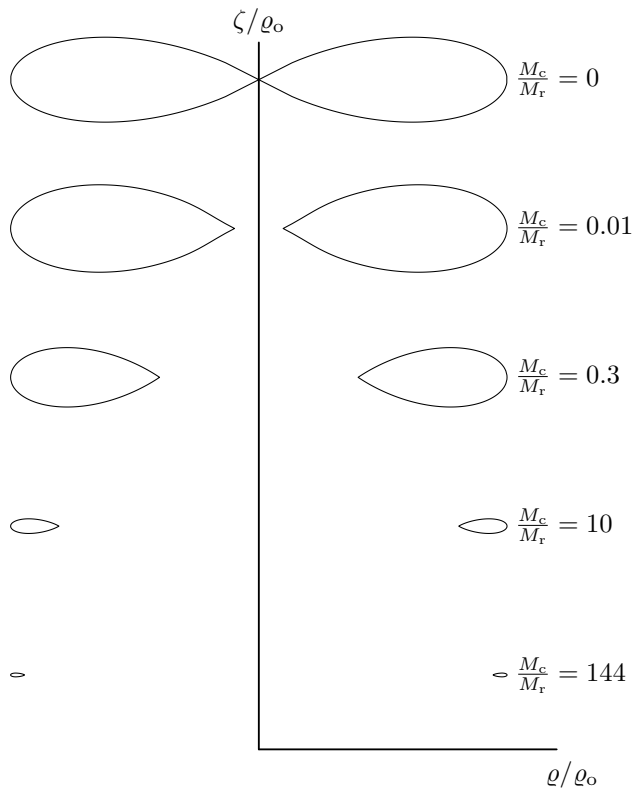}
 \caption{Cross-sections of Newtonian rings surrounding a point mass with varying ratios of central to ring mass $M_{\text c}/M_{\text r}$. The normalized coordinate $\zeta/\varrho_{\text o}$ is plotted agianst $\varrho/\varrho_{\text o}$. Each of these configurations possesses an inner mass-shed and has a normalized total mass of  $M_{\text{tot}}\sqrt{\varepsilon}=(M_{\text c}+M_{\text r})\sqrt{\varepsilon}=1$. \label{mass-shed}}
\end{figure}

If we consider the sequence of configurations at the inner mass-shedding limit and with constant total mass, then we can vary a third parameter such as $M_{\text c}/M_{\text r}$. In the limit for which this ratio of masses goes to zero (i.e.\ when the point mass vanishes), we arrive at the configuration denoted by `(H)' in Fig.~6 of \cite{AKM2}. As described there, such a configuration can be found along the sequence bifurcating from the Maclaurin spheroid with an eccentricity of $\epsilon=0.98523\ldots$ and marks the transition from a spheroidal to a toroiodal topology. Presumably, there is no upper limit to the value of $M_{\text c}/M_{\text r}$ that can be reached. However, this test-ring limit could only be reached if the ring were not of finite size, i.e.\ in the limit $M_{\text c}/M_{\text r} \to \infty$ it follows that $\varrho_{\text i}/\varrho_{\text o} \to 1$. The cross-sections for configurations with an inner mass-shed can be seen in Fig.~\ref{mass-shed}.

\subsection{A Ring Surrounding a Black Hole}

\begin{figure}
 \includegraphics{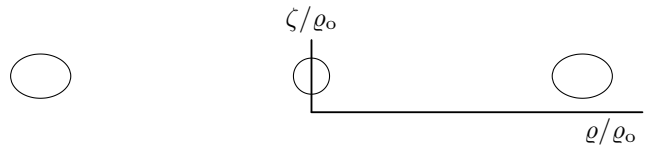}
 \caption{The cross-section of the ring surrounding the Black Hole described in Table~\ref{tab:J0}. The normalized coordinate $\zeta/\varrho_{\text o}$ is plotted against $\varrho/\varrho_{\text o}$. Note that in these coordinates, the Black Hole always appears as a circle. \label{cross_J0}}
\end{figure}

An example of the convergence and extreme accuracy of the numerical relativistic program can be found in Table~\ref{tab:J0}. All of the dimensionless parameters listed in this section (and denoted by a bar) are normalized with respect to the (constant) energy density $\varepsilon$. Thus we have
\begin{alignat*}{3}
 \bar{M}_{\text{tot}} &:= M_{\text{tot}}\,\sqrt{\varepsilon} & \quad
 \bar{M}_{\text{c}} &:= M_{\text{c}}\,\sqrt{\varepsilon} & \quad
 \bar{M}_{\text{r}} &:= M_{\text{r}}\,\sqrt{\varepsilon} \\
 \bar{J}_{\text{tot}} &:= J_{\text{tot}}\, \varepsilon & \quad
 \bar{J}_{\text{c}} &:= J_{\text{c}}\, \varepsilon & \quad
 \bar{J}_{\text{r}} &:= J_{\text{r}}\, \varepsilon \\
 \bar{\Omega}_{\text{c}} &:= \Omega_{\text{c}}/\sqrt{\varepsilon} & \quad
 \bar{\Omega}_{\text{r}} &:= \Omega_{\text{r}}/\sqrt{\varepsilon} & \quad
 \bar{\varrho}_{\text o}       &:= \varrho_{\text o}\,\sqrt{\varepsilon}.
 \end{alignat*} 
The configuration in Table~\ref{tab:J0} was calculated by prescribing $J_{\text c}/\varrho_{\text o}^{\,2}=0$, $\varrho_{\text i}/\varrho_{\text o}=0.8$,  $M_{\text{tot}}/\varrho_{\text o}=0.24$ and $r_{\text c}/\varrho_{\text o}=0.06$. An indication of the accuracy of the solution at each approximation order can be found in the last two rows, in which the total mass and angular momentum are calculated via eq.~\eqref{asymptotics}
\[M_\text o:= M_\text{tot} \quad  J_\text o:=J_\text{tot} \qquad \text{calculated at infinity}\]
and compared to the values
\[M_{\text i} := M_{\text{tot}} \quad J_{\text i} := J_{\text{tot}} \qquad \text{calculated via integral}\]
found by adding eq.~\eqref{int_Mring} to \eqref{MBH} for the mass and eq.~\eqref{int_Jring} to \eqref{JBH} for the angular momentum. All the digits listed in the final column are valid, showing that machine accuracy can be reached. With such high accuracy it is possible to study effects that were misunderstood previously due to the slight inaccuracies associated with older numerical methods. In particular, Nishida \& Eriguchi \cite{NE94} presumed that $R_{\text p}/R_{\text e}=1$ is strictly valid for $J_{\text c}=0$, i.e.\ that a Black Hole with no angular momentum has a non-deformed horizon. One can see in the seventh row of Table~\ref{tab:J0}, that the value is indeed very close to one, but not strictly equal to one. A coordinate cross-section of the surface of the ring and the horizon of the Black Hole can be found in Fig.~\ref{cross_J0}.

\begin{table*}
 \begin{tabular}{ccccc} \hline
       m                      & 8         &      16    &     22       &       28 \\ \hline 
  $\bar{M}_{\text{tot}}$    & 0.137611  & 0.13760306 & 0.1376030595 & 0.137603059537\\    %
  $\bar{M}_{\text{c}}$      & 0.0687121 & 0.068707899 & 0.06870789942 & 0.0687078994344\\ %
  $\bar{M}_{\text{r}}$      & 0.0689223 & 0.068895164 & 0.06889516011 & 0.068895160102\\ 
  $\bar{J}_{\text{tot}}$    & 0.0203957 & 0.020389051 & 0.02038905076 & 0.0203890507611\\ %
  $\bar{\Omega}_{\text{c}}$ & 0.133435  & 0.13342471 & 0.1334247047 & 0.133424704700\\ %
  $\bar{\Omega}_{\text{r}}$ & 0.603778  & 0.60387735 & 0.6038773604 & 0.603877360278\\ %
  $R_{\text p}/R_{\text e}$ & 0.998385  & 0.99833254 & 0.9983324786 & 0.99833247818 \\
  $\bar \varrho_{\text o}$        & 0.573379  & 0.57334607 & 0.5733460813 & 0.57334608140 \\ \hline %
  $\left|\frac{M_{\text i}-M_{\text o}}{M_{\text o}}\right|$
             & $2 \times 10^{-4}$ & 4 $\times 10^{-8}$ & $1 \times 10^{-10}$ & $1 \times 10^{-13}$ \\
  $\left|\frac{J_{\text i}-J_{\text o}}{J_{\text o}}\right|$ 
             & $2 \times 10^{-5}$ & $1 \times 10^{-8}$ & $5 \times 10^{-11}$ & $2 \times 10^{-13}$ \\ \hline
 \end{tabular}
 \caption{Physical parameters for a ring surrounding a Black Hole showing the convergence of the numerical solution for increasing order, $m$, of the polynomial approximation. The configuration was determined by prescribing $J_{\text c}/\varrho_{\text o}^{\,2}=0$, $\varrho_{\text i}/\varrho_{\text o}=0.8$,  $M_{\text{tot}}/\varrho_{\text o}=0.24$ and $r_{\text c}/\varrho_{\text o}=0.06$. \label{tab:J0}}
\end{table*}

\begin{figure}
 \includegraphics[width=\columnwidth]{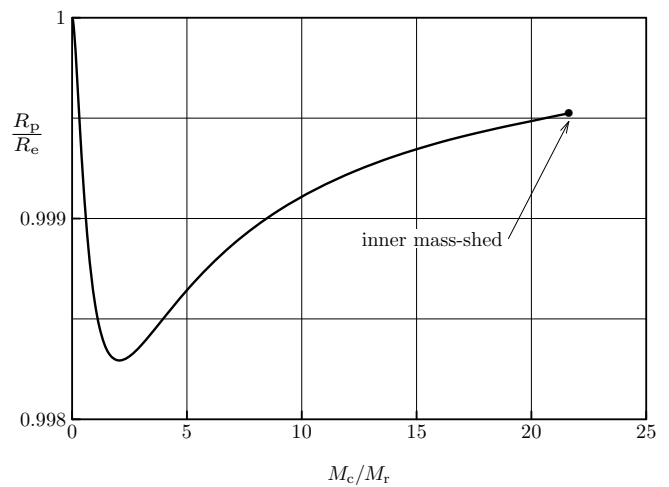}
 \caption{$R_{\text p}/R_{\text e}$ versus $M_{\text c}/M_{\text r}$ is plotted for configurations with $J_{\text c}/\varrho_{\text o}^{\,2}=0$, $\varrho_{\text i}/\varrho_{\text o}=0.9$ and $M_{\text{tot}}/\varrho_{\text o}=0.24$. Despite the fact that the Black Hole has no angular momentum, a small deformation of the horizon is apparent. \label{fig:J0}}
\end{figure}

\begin{figure}
 \includegraphics[width=\columnwidth]{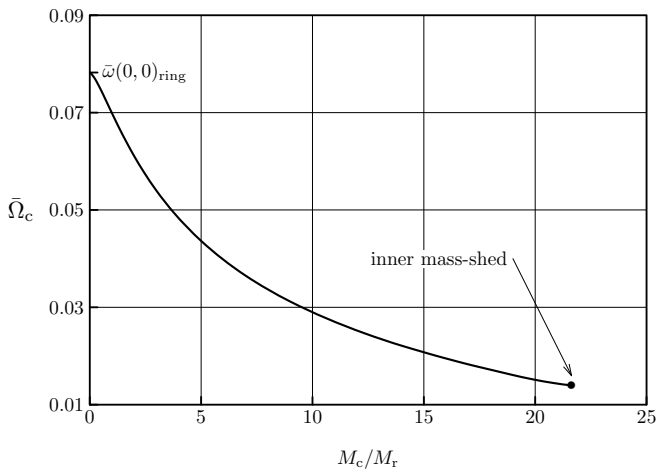}
 \caption{$\bar \Omega_{\text c}$ versus $M_{\text c}/M_{\text r}$ is plotted for the configurations of Fig.~\ref{fig:J0}, i.e.\ with $J_{\text c}/\varrho_{\text o}^{\,2}=0$, $\varrho_{\text i}/\varrho_{\text o}=0.9$ and $M_{\text{tot}}/\varrho_{\text o}=0.24$. The label $\bar\omega(0,0)_{\text{ring}}$ indicates the value that the metric function $\bar \omega=\omega/\sqrt{\varepsilon}$ assumes at the point $(\varrho,\zeta)=(0,0)$ when the Black Hole is absent. \label{fig:J0_Om}}
\end{figure}

We calculated a series of configurations for which we prescribed the values $J_{\text c}/\varrho_\text o^{\,2}=0$, $\varrho_{\text i}/\varrho_{\text o}=0.9$ and $M_{\text{tot}}/\varrho_{\text o}=0.24$ and varied the ratio of the Black Hole's to the ring's mass. In Fig.~\ref{fig:J0} we plot $R_{\text p}/R_{\text e}$ versus $M_{\text c}/M_{\text r}$ for the entirety of this sequence, i.e.\ from the limit $M_{\text c}/M_{\text r} \to 0$ right up to an endpoint, which turned out to be an inner-mass shedding limit (and hence analogous to the Newtonian results). Because of the exponential coordinates that were discussed in \S~\ref{methods}, it was possible to calculate configurations with $M_{\text c}/M_{\text r}$ very close to zero (e.g.\ $M_{\text c}/M_{\text r} \approx 1/50$). We see that as the mass of the Black Hole tends to zero, the deformation of its horizon vanishes. This may well be because it shrinks to a point in this limit and no tidal forces are present to distort it. In essence, this is the limit of an infinitesimal spherical Black Hole at the origin of the spacetime in the external field of the ring. Fig.~\ref{fig:J0_Om} lends credit to this interpretation. For the same sequence of configurations, we have plotted $\bar \Omega_{\text c}$ versus $M_{\text c}/M_{\text r}$ and indicated the value $\omega(\varrho=0,\zeta=0)/ \sqrt{\varepsilon}= \bar\omega(0,0)_{\text{ring}}= 0.07823\ldots$ that this metric function assumes at the point $(\varrho,\zeta)=(0,0)$ in the absence of a Black Hole (as calculated with a numerical program as described in \cite{AKM4}). In the limit $M_{\text c}/M_{\text r} \to 0$, the Black Hole, which has no angular momentum, is not flattened, although it rotates with precisely the angular velocity that arises due to the frame dragging effect of the ring.

It should be noted that whereas Figs~\ref{fig:J0} and \ref{fig:J0_Om} would likely look qualitatively similar for different values of $\varrho_{\text i}/\varrho_{\text o}$ and $M_{\text{tot}}/\varrho_{\text o}$, the choice $J_{\text c}/\varrho_{\text o}^{\,2}=0$ is special. Had we chosen a non-zero value for $J_{\text c}/\varrho_{\text o}^{\,2}$, then it is unlikely that we could have reached an arbitrarily small value for $M_{\text c}/\varrho_{\text o}$ since $J_{\text c}/M_{\text c}^{\,2}$ cannot, most likely, become arbitrarily large. In other words, we expect that a physical limit, analogous to that reached by the extreme Kerr solution, will prohibit the possibility of prescribing arbitrarily large values for $J_{\text c}/M_{\text c}^{\,2}$. We shall see, however, toward the end of this section that it can become greater than one. Moreover, by fixing a value for $M_{\text{tot}}/\varrho_{\text o}$, we have precluded the possibility that $M_{\text r}/\varrho_{\text o}$ can become arbitrarily large so that the limit $M_{\text c}/M_{\text r} \to 0$ would not exist. Fixing a value for $\varrho_{\text i}/\varrho_{\text o}$, we also precluded the limit $M_{\text c}/M_{\text r} \to \infty$, since the sequence would end in a mass-shedding limit.

\begin{figure}
 \includegraphics[width=\columnwidth]{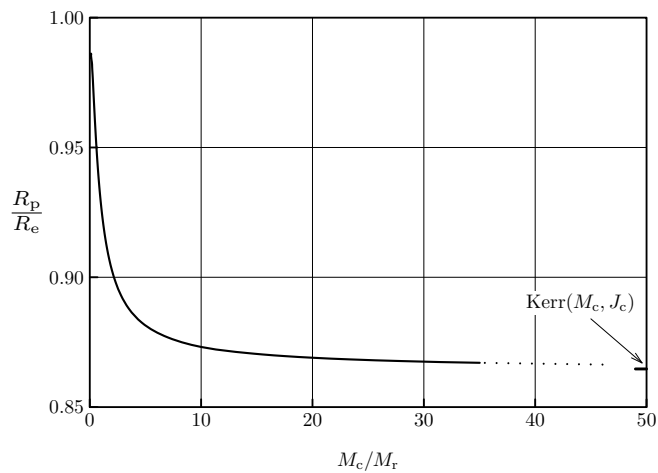}
 \caption{$R_{\text p}/R_{\text e}$ versus $M_{\text c}/M_{\text r}$ is plotted for configurations with $M_{\text c}/\varrho_{\text o}=0.14$, $J_{\text c}/\varrho_{\text o}^{\,2}=0.015$ and $\varepsilon\,\varrho_{\text o}^{\,2}=0.24$. The dotted lines convey that this sequence continues on and tends to the value indicated by $\text{Kerr}(M_{\text c},J_{\text c})$. The label $\text{Kerr}(M_{\text c},J_{\text c})$ indicates the value for $R_{\text p}/R_{\text e}$ for a Kerr Black Hole with the same mass and angular momentum as were prescribed here.\label{fig:Kerr}}
\end{figure}

\begin{figure}
 \includegraphics[width=\columnwidth]{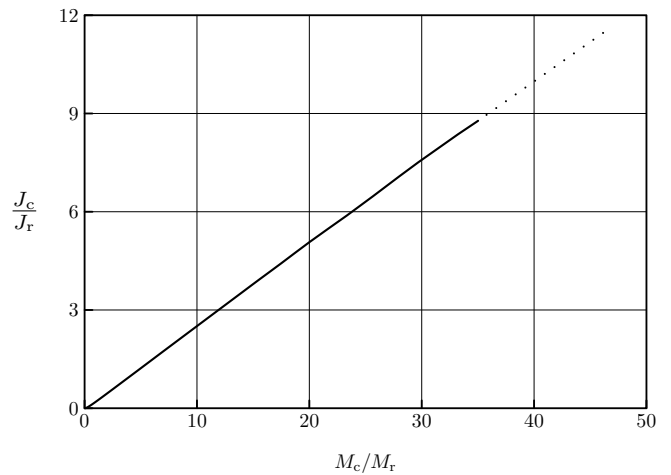}
 \caption{$J_{\text c}/J_{\text r}$ versus $M_{\text c}/M_{\text r}$ is plotted for the series of configurations in Fig.~\ref{fig:Kerr}. As in that figure, the dotted lines convey that this sequence continues on.\label{fig:Kerr_J}}
\end{figure}

In Fig.~\ref{fig:Kerr} we provide an example of a significantly distorted horizon. By prescribing the angular momentum and mass of the Black Hole, we are able to make use of eq.~\eqref{eq:Kerr_RR} and compare values when a ring is present to those when it is absent. We plot $R_{\text p}/R_{\text e}$ versus $M_{\text c}/M_{\text r}$ for a series of configurations for which we prescribe $M_{\text c}/\varrho_{\text o}=0.14$, $J_{\text c}/\varrho_{\text o}^{\,2}=0.015$ and $\varepsilon\,\varrho_{\text o}^{\,2}=0.24$, but allow $M_{\text r}/\varrho_{\text o}$ to become arbitrarily large and can thus approach the limit $M_{\text c}/M_{\text r} \to 0$. As the mass of the ring tends to zero (i.e.\ $M_{\text c}/M_{\text r} \to \infty$), we see that the distortion of the horizon approaches that of a Kerr Black Hole. As one increases the mass of the ring however, the deviation from this oblateness becomes significant. Of interest is the fact that, in contrast to Fig.~\ref{fig:J0}, the value of $R_{\text p}/R_{\text e}$ does not approach its unperturbed value as the relative mass of the Black Hole becomes negligible. This may be related to the fact that here the angular momentum of the ring as well as its mass tend to infinity. This is illustrated in Fig.~\ref{fig:Kerr_J} in which $J_{\text c}/J_{\text r}$ is plotted against $M_{\text c}/M_{\text r}$ and the curve intersects the origin. The sequence plotted here was cut off at an arbitrary value for $M_{\text c}/M_{\text r}$ as indicated by the dotted lines. The two limits that are relevant to Figs~\ref{fig:Kerr} and \ref{fig:Kerr_J}, $M_{\text c}/M_{\text r} \to 0$ and $M_{\text c}/M_{\text r} \to \infty$, are also interesting in that they represent two entirely different limits for a system containing an infinitely thin ring. A more thorough investigation of such limits is planned for a future publication.

\begin{figure}
 \includegraphics{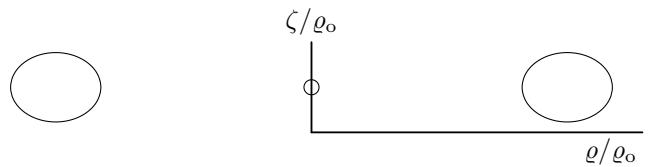}
 \caption{The cross-section of the ring surrounding the Black Hole described in Table~\ref{tab:M2J}. The normalized coordinate $\zeta/\varrho_{\text o}$ is plotted against $\varrho/\varrho_{\text o}$. Note that in these coordinates, the Black Hole always appears as a circle. \label{cross_M2J}}
\end{figure}

Finally, we provide a second table showing the convergence and high accuracy of the program. What is particularly interesting in Table~\ref{tab:M2J} is that the value of the angular momentum of the Black Hole divided by the square of its mass exceeds one. We prescribed the values $J_{\text c}/M_{\text c}^{\, 2}=20/19$, $\varrho_{\text i}/\varrho_{\text o}=0.7$,  $M/\varrho_{\text o}=0.35$ and $r_{\text c}/\varrho_{\text o}=0.025$ and were able to reach machine accuracy. The horizon of this Black Hole and surface of the ring in cross-section  can be seen in Fig.~\ref{cross_M2J}.

\begin{table*}
 \begin{tabular}{ccccc} \hline
       m                      & 8          & 16           &        22       &       28 \\ \hline 
  $\bar{M}_{\text{tot}}$    & 0.177955   & 0.17795018   & 0.1779501819    & 0.177950181905\\ 
  $\bar{M}_{\text{c}}$      & 0.0314095  & 0.031405521  & 0.03140551959   & 0.031405519584\\ 
  $\bar{M}_{\text{r}}$      & 0.146562   & 0.14654468   & 0.1465446624    & 0.14654466232\\ 
  $\bar{J}_{\text{tot}}$    & 0.0407783  & 0.040765569  & 0.04076556804   & 0.040765568044\\ 
  $\bar{J}_{\text{c}}$      & 0.00103848 & 0.0010382176 & 0.001038217538  & 0.0010382175372 \\
  $\bar{J}_{\text{r}}$      & 0.0397165  & 0.039727351  & 0.03972735051   & 0.039727350507\\
  $\bar{\Omega}_{\text{c}}$ & 2.99321    & 2.9924299    & 2.992429509     & 2.9924295058\\ 
  $\bar{\Omega}_{\text{r}}$ & 0.656494   & 0.65661265   & 0.6566127086    & 0.65661270903\\ 
  $R_{\text p}/R_{\text e}$ & 0.930597   & 0.93038725   & 0.9303867661    & 0.9303867600\\
  $\bar \varrho_{\text o}$        & 0.508443   & 0.50842910   & 0.5084290912    & 0.50842909116  \\ \hline
  $\left|\frac{M_{\text i}-M_{\text o}}{M_{\text o}}\right|$
             & $9 \times 10^{-5}$ & $1 \times 10^{-7}$ & $7 \times 10^{-10}$ & $7 \times 10^{-12}$\\
  $\left|\frac{J_{\text i}-J_{\text o}}{J_{\text o}}\right|$ 
             & $6 \times 10^{-4}$ & $1 \times 10^{-8}$ & $7 \times 10^{-11}$ & $5 \times 10^{-13}$\\ \hline
 \end{tabular}
 \caption{Physical parameters for a ring surrounding a Black Hole showing the convergence of the numerical solution for increasing order of the polynomial approximation. The configuration was determined by prescribing $J_{\text c}/M_{\text c}^{\,2}=20/19$, $\varrho_{\text i}/\varrho_{\text o}=0.7$,  $M_\text{tot}/\varrho_{\text o}=0.35$ and $r_{\text c}/\varrho_{\text o}=0.025$. \label{tab:M2J}}
\end{table*}

%% file: Future.tex
We have seen that it is possible to calculate axisymmetric, stationary configurations consisting of a Black Hole surrounded by a ring of matter numerically up to machine accuracy. The basic ideas behind the numerical methods are not all that different from those presented in \cite{AKM3}, but there are specific numerical challenges that must be overcome and which were, in part, presented here. Such a numerical code allowed us to take a look at the influence of matter on the properties of the Black Hole and we saw in Figs~\ref{fig:J0} and \ref{fig:Kerr} how the shape of the horizon deviates from its unperturbed value. We also saw that the presence of the ring allows us to construct situations in which $J_\text c/M_\text c^{\,2}>1$ for the angular momentum and mass of the Black Hole.

The fact that the configurations considered here contain four parameters means that a rigorous exploration of the solution space is an ambitious task. We thus first intend to focus our attention on particular aspects of the solutions that we believe could prove fruitful.
These include: (1) analysing the limits that hold for $J_{\text c}/M_{\text c}^{\,2}$, (2) studying the influence of matter on the Black Hole more extensively by considering multipole moments on the horizon (see e.g.\ \cite{AEPB2004}), (3) considering other equations of state and (4) exploring a possible parametric transition to an infinitely flattened ring.